# Polarization transformations by a magneto-photonic layered structure in vicinity of ferromagnetic resonance

**Vladimir R Tuz**[1,2], **Mikhail Yu Vidil**[2], **and Sergey L Prosvirnin**[1,2]

[1]School of Radio Physics, Karazin Kharkiv National University, 4, Svobody Square, Kharkiv 61077, Ukraine

[2]Institute of Radio Astronomy of National Academy of Sciences of Ukraine, 4, Krasnoznamennaya st., Kharkiv 61002, Ukraine

E-mail: Vladimir.R.Tuz@univer.kharkov.ua, vidil@rian.kharkov.ua and prosvirn@rian.kharkov.ua

**Abstract**

Polarization properties of a magnetophotonic layered structure are studied at the frequencies close to the frequency of ferromagnetic resonance. The investigations are carried out taking into account a great value of dissipative losses in biased ferrite layers in this frequency band. The method is based on analysis of solution stability of ordinary differential equation system regarding the field inside a periodic stack of dielectric and ferrite layers. The electromagnetic properties of the structure are ascertained by the analysis of the eigenvalues of the transfer matrix of the structure period. The frequency boundaries of the stopbands and passbands of the eigenwaves are determined. The frequency and angular dependences of the reflection and transmission coefficients of the stack are presented. Frequency dependences of a polarization rotation angle and ellipticity of the reflected and transmitted fields are analyzed. An enhancement of polarization rotation due to the periodic stack is mentioned, in comparison with the rotation due to some effective ferrite slab.

(Some figures in this article are in color only in the electronic version)

## 1. Introduction

Electromagnetic properties of materials that have artificially created periodic translation symmetry differ significantly from those of the homogeneous media. There is a direct analogy between the wave processes in such structures and the properties of the wave functions of electrons moving in the periodic potential of a crystal lattice. The translation symmetry significantly affects the spectrum of eigenwaves of such materials. There are alternating bandwidths in which propagation of electromagnetic waves is either possible (passbands) or forbidden (stopbands). Due to these properties, lately, it is conventional to refer such one-, two- or three-dimensional periodic structures composed of substantial index contrast elements to photonic crystals (PCs). The forbidden bandwidths are called photonic band gaps (PBG) [1, 2]. Note that in the one-dimensional case a PC is nothing more than a dielectric periodic layered structure. We use this notation trough this paper to emphasize the contrast of such structures to 2D or 3D periodic PC [3]. The PCs are now widely used in modern integrated optics and optoelectronics, laser and X-ray techniques, microwave and optical communications.

From the viewpoint of applications, it is obvious that not only design of PCs but also control of the position and width of the band gap is of a great interest. One of the ways to realize the control is using magnetic materials in fabrication of PCs to produce so-called magneto-photonic crystals (MPCs) [4-6]. Indeed, a biased external static magnetic field can alter permittivity or permeability of MPC ingredients. In addition to the possibility to control properties of the PCs, the MPCs manifest some unique magneto-optical properties accompanied by the Kerr and Faraday rotation enhancement arising from the effect of light localization as a result of wave interference within the magnetic structure. Owing to these features, MPCs have already found several electronic applications as isolators and circulators and magneto-optic spatial light modulators. We call these multilayered one-dimensional structures as magneto-photonic layered structures (MPLS), unlike 2D or 3D periodic MPCs.

According to classical electromagnetics, the ferromagnetic properties of media are related to a spin magnetic moment [7, 8]. In the presence of an external static magnetic field, the electron spin precess around the field with a frequency $\omega_0$, that is called as the frequency of ferromagnetic resonance. The phenomenon of the ferromagnetic resonance plays a crucial role, because it determines largely the magnitude of dissipative losses, Faraday rotation and the nonreciprocity of medium. Therefore, the properties of a MPLS depend strongly on the proximity of an incident wave frequency to the frequency of the ferromagnetic resonance. Thus a wave propagating through MPLS experiences its attenuation depending on the ratio of the wave frequency $\omega$ to the resonant frequency $\omega_0$. In particular, the wave decay will reach maximum at the frequency $\omega = \omega_0$. In spite of the fact that the magnetic losses depend on the frequency, in most studies of MPLSs the frequency of a propagating wave is chosen quite far from the resonance frequency $\omega_0$, i.e. a special situation of infinitesimally small dissipative losses is under consideration [4-6].

In addition to a strong dissipative attenuation, the resonance magnetic losses affect also the polarization characteristics of MPLS. For an observer looking in the direction of the biased magnetic field, electron spins precess clockwise [7, 8]. Thus precessing electron spins have different effect on the right-handed and left-handed circularly polarized wave propagated along the biased field. Through the paper we use the optical definition of a circular polarization. We assume the wave has the right-handed (left-handed) circular polarization, if its vector of electric field rotates clockwise (anticlockwise) to an observer looking opposite to the wave propagation direction. Only for a left-handed circularly polarized wave propagated along biased magnetic field, the resonance phenomenon occurs causing the difference of effective complex magnetic permeability related to the right-handed and left-handed polarized waves. It yields the magnetic rotation of the polarization plane of the linearly polarized wave, which has great practical interest.

In most papers, the normal wave incidence is considered and the transversal or longitudinal magneto-optic configuration of the biased external static magnetic field is chosen to study the electromagnetic properties of MPLSs. In the transversal biased field configuration, the electromagnetic wave can be presented as $TE-$ and $TM-$waves [7, 10-12, 14], and in the longitudinal one, as the right-handed and left-handed circularly polarized waves [8, 10, 11]. In either case these modes are uncoupled ones and the solution of electromagnetic wave propagation problem is described via a $2 \times 2$ transfer matrix formulation. Generally when a MPLS consists of ferromagnetic layers with arbitrary orientation of the anisotropy axes or the wave impinges obliquely, the modes are right-handed and left-handed elliptically polarized, and it needs to use a $4 \times 4$ transfer matrix [6, 13, 15-17].

The essence of the transfer matrix method consists in derivation of the matrix to relate the tangential field components at the beginning and the end of the structure period and thereafter treatment of this matrix to study both eigenwaves and fields reflected from and transmitted through MPLS. As it is well known, the transfer matrix of the period of the layered structure is a product of the transfer matrices of its separate layers. After the transfer matrix of the period is found, further investigation is based on the analysis of the eigenvalues of this matrix since they are directly related to the properties of eigenwaves of the periodic structure.

By definition, the eigenwaves of a periodic structure is non-trivial solutions of the homogeneous Maxwell equations that satisfy the conditions of the Floquet's quasi-periodicity [18]

$$\vec{\Psi}(x,y,z) = \exp(i\gamma L)\vec{\Psi}(x,y,z+L),  \qquad (1)$$

where $\vec{\Psi}$ is some linear function of the field components, $L$ is the structure period. The condition (1) expresses the intuitive notion about waves in a periodic structure: in the neighboring periods the field differs only by a certain phase factor $\gamma$. This parameter $\gamma$ is a wavenumber related to a certain eigenwave. They are named as the Bloch wavenumber and the Bloch wave, respectively. It is obvious that the Bloch wave exists in any section of an infinite periodic medium. For lossless structures, the parameter $\gamma$ takes either purely real or purely imaginary values in the passbands and stopbands, respectively.

When the material losses are taken into consideration, the formal solution of the dispersion equation leads to complex values of $\gamma$ ($\gamma = \gamma' + i\gamma''$). In this case an exponential decay of the field follows from the condition (1), which gives a contradiction to the definition of the Bloch waves in an infinite periodic structure. Thus it is necessary to initially refuse the imposition of the condition (1). Instead, the standard method of the theory of irregular waveguides [18] can be applied. In the context of this theory it is assumed that the eigenwaves of the irregular waveguide with impedance sidewalls are orthogonal in energy terms. It means that every eigenwave propagates independently from the others inside the area free of sources. Thus the eigenwaves have clear physical meaning: it is the field that can be excited in the waveguide outside the area occupied by the sources. The method is based on obtaining a system of ordinary differential equations and further analysis of the stability of solutions of this system. In simple terms, the concept of stability is associated with the analysis of behavior of small deviations from the trivial solution of the differential equation [19]. In the case of periodic media, the corresponding system of equations contains periodic coefficients, and, from a mathematical point of view, the transfer matrix defines the fundamental solutions of the system on a dedicated interval. In particular, as in our present work, the electromagnetic properties of the structure under study are found by the analysis of the eigenvalues of the transfer matrix, as these eigenvalues describe the stability of solutions of the system.

In the present paper, we focus on the study of electromagnetic properties of a finite one-dimensional periodic layered structure that consists of normally biased ferromagnetic and isotropic layers. A distinctive feature of this work is a study of the properties of MPLS in a frequency range of the ferromagnetic resonance. In particular the polarization transformations are under investigation. Our goal is to derive the solution of the problem of electromagnetic wave propagation in the MPLS with dissipative losses typical over the frequency range close to the frequency of ferromagnetic resonance. In our theory treatment, we deny a prior choice of solution in the common form (1) and employ the general theory of differential equations in order to include other possible forms of solution into consideration. We also study peculiarities of field transformations by MPLS in the case of oblique incidence of electromagnetic wave.

**2. Problem formulation**

A periodic stack of $N$ identical double-layer slabs placed along $z$-axis is investigated, see Fig. 1. We assume that the structure is isotropic and infinite in the $x$- and $y$- directions. Each slab consists of a ferrite layer and a homogeneous isotropic dielectric layer. The ferrite layer is magnetized up to saturation by an external static magnetic field $\vec{H}_0$ directed normally to the layer. Constitutive parameters of ferrite layer are $\varepsilon_1$, $\hat{\mu}_1$ and its thickness is $d_1$. The dielectric layer has a thickness $d_2$, permittivity $\varepsilon_2$, and permeability $\mu_2$. Thus the thickness of the slab of periodic structure is $L = d_1 + d_2$.

The outer half-spaces $z \leq 0$ and $z \geq NL$ are homogeneous, isotropic and have constitutive parameters $\varepsilon_0$, $\mu_0$ and $\varepsilon_3$, $\mu_3$, respectively. Let us suppose that an incident field is a plane wave of a frequency $\omega$ and its direction of propagation, in the region $z \leq 0$, is determined by angles $\theta_0$ and $\varphi_0$ relative to $z$-axis and $x$-axis respectively (see Fig. 1). Time dependence is assumed $\exp(-i\omega t)$ through the paper.

We use common expressions for constitutive parameters of normally biased ferrite taking into account the magnetic losses [20]

$$\varepsilon_1 = \varepsilon_f, \qquad \hat{\mu}_1 = \begin{pmatrix} \mu_1^T & i\alpha & 0 \\ -i\alpha & \mu_1^T & 0 \\ 0 & 0 & \mu_1^L \end{pmatrix}, \tag{2}$$

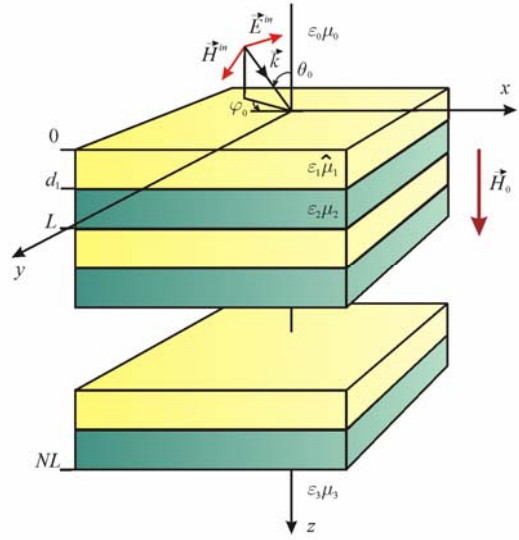

**Figure 1. (color online)** A periodic stack of double-layer slabs composed of normally magnetized ferrite and isotropic dielectric layers

where $\mu_1^T = 1 + \chi' + i\chi''$, $\chi' = \omega_0\omega_m\left[\omega_0^2 - \omega^2(1-b^2)\right]D^{-1}$, $\chi'' = \omega\omega_m b\left[\omega_0^2 + \omega^2(1+b^2)\right]D^{-1}$, $\alpha = \Omega' + i\Omega''$, $\Omega' = \omega\omega_m\left[\omega_0^2 - \omega^2(1+b^2)\right]D^{-1}$, $\Omega'' = 2\omega^2\omega_0\omega_m bD^{-1}$, $D = \left[\omega_0^2 - \omega^2(1+b^2)\right]^2 + 4\omega_0^2\omega^2 b^2$, and typical parameters in the microwave region are $\varepsilon_f = 10$, $\mu_1^L = 1$, $b = 0.05$, $\omega_0/2\pi = 4\text{GHz}$, $\omega_m/2\pi = 5.6\text{GHz}$. The value $\omega_m$ corresponds to saturation magnetization of 2000 G [20]. The frequency dependences of the permeability parameters are presented in Fig. 2. Note, the values of $\text{Im}\,\mu^T$ and $\text{Im}\,\alpha$ are so close to each other that the curves of their frequency dependences coincide in the figure.

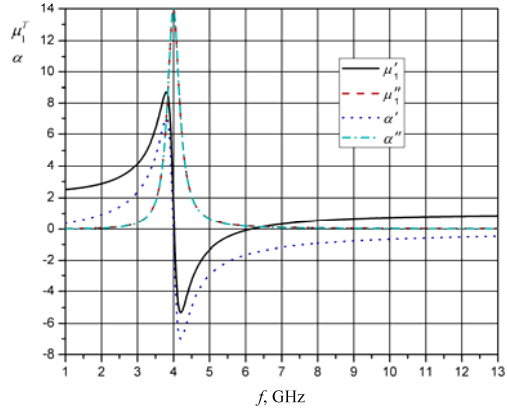

**Figure 2. (color online)** Frequency dependences of permeability parameters of along to $z$-axis biased ferrite layer.

### 3. Formalism of $4\times 4$ transfer matrix

Let us consider the incidence of a plane monochromatic wave on a layered structure in which permittivity $\varepsilon$ and permeability $\hat{\mu}$ are scalar and tensor piecewise constant functions of the coordinate $z$ respectively. In Cartesian coordinates the system of Maxwell's equations *for each layer* has a form

$$ik_y H_z - \frac{\partial H_y}{\partial z} = -ik\varepsilon E_x, \quad ik_y E_z - \frac{\partial E_y}{\partial z} = ik(\hat{\mu}\vec{H})_x,$$

$$\frac{\partial H_x}{\partial z} - ik_x H_z = -ik\varepsilon E_y, \quad \frac{\partial E_x}{\partial z} - ik_y E_z = ik(\hat{\mu}\vec{H})_y, \qquad (3)$$

$$ik_x H_y - ik_y H_x = -ik\varepsilon E_z, \quad ik_x E_y - ik_y E_x = ik(\hat{\mu}\vec{H})_z.$$

where $k_x = k\sin\theta_0 \cos\varphi_0$, $k_y = k\sin\theta_0 \sin\varphi_0$, $k = \omega/c$ is a free-space wave number. From six components of the electromagnetic field $\vec{E}$ and $\vec{H}$, only four ones are independent. Thus we can eliminate the components $E_z$ and $H_z$ from the system (3) and derive a set of four first-order linear differential equations in the transversal field components inside a layer of the structure

$$\frac{\partial}{\partial z}\begin{pmatrix} E_x \\ E_y \\ H_x \\ H_y \end{pmatrix} = ik \begin{pmatrix} 0 & 0 & i\alpha + k_x k_y/\varepsilon k^2 & \mu^T - k_x^2/\varepsilon k^2 \\ 0 & 0 & -\mu^T + k_y^2/\varepsilon k^2 & i\alpha - k_x k_y/\varepsilon k^2 \\ -k_x k_y/\mu^L k^2 & -\varepsilon + k_x^2/\mu^L k^2 & 0 & 0 \\ \varepsilon - k_y^2/\mu^L k^2 & k_x k_y/\mu^L k^2 & 0 & 0 \end{pmatrix}\begin{pmatrix} E_x \\ E_y \\ H_x \\ H_y \end{pmatrix}, \qquad (4)$$

where $\varepsilon = \varepsilon_1$, $\alpha \neq 0$, $\mu^T = \mu_1^T$, $\mu^L = \mu_1^L$ in the ferrite layers $mL \leq z \leq mL + d_1$ and $\varepsilon = \varepsilon_2$, $\alpha = 0$, $\mu^T = \mu^L = \mu_2$ in the dielectric layers $mL + d_1 \leq z \leq (m+1)L$, $m = 0,1,...,N-1$.

The set of equations (4) can be abbreviated by using a matrix formulation

$$\frac{\partial}{\partial z}\mathbf{\Psi}(z) = ik\mathbf{A}(z)\mathbf{\Psi}(z), \qquad (5)$$

where $\mathbf{A}(z)$ is a $4\times 4$ matrix, and $\mathbf{\Psi}$ is a four-component vector with evident notations.

Further, we assume that the vector $\mathbf{\Psi}(z)$ is known in the plane $z = z_0$ and find a solution of the *Cauchy problem* for equation (5) in the form

$$\mathbf{\Psi}(z) = \mathbf{M}(z, z_0)\mathbf{\Psi}(z_0). \qquad (6)$$

In electromagnetic theory the matrix $\mathbf{M}(z, z_0)$ is referred to as a *transfer* matrix. The matrix $\mathbf{M}(z, z_0)$ satisfies the following relations [19]

$$\mathbf{M}(z_0, z_0) = \mathbf{I},$$
$$\mathbf{M}(z, z_1)\mathbf{M}(z_1, z_2) = \mathbf{M}(z, z_2), \qquad (7)$$
$$\mathbf{M}^{-1}(z_1, z_2) = \mathbf{M}(z_2, z_1),$$

and

$$\mathbf{M}(z, z_0) = \mathbf{M}_0(z)\mathbf{M}_0^{-1}(z_0), \qquad (8)$$

where $\mathbf{M}_0(z) \equiv \mathbf{M}(z, 0)$, and $\mathbf{I}$ is the $4\times 4$ identity matrix.

Let us consider the case of a periodical coefficient of the system (5), i.e.
$$\mathbf{A}(z + L) = \mathbf{A}(z) \qquad (9)$$

or in terms of the transfer matrix
$$\mathbf{M}(z + L, z_0 + L) = \mathbf{M}(z, z_0) \qquad (10)$$

where $L$ is period.

Taking into account (7) and (8), a multiplicative identity is derived from (10)
$$\mathbf{M}_0(z + L) = \mathbf{M}_0(z)\mathbf{M}_0(L). \qquad (11)$$

The constant matrix $\mathbf{M}_0(L)$ is called as the *monodromy* matrix [19]. Further, if it is assumed that
$$\mathbf{K} \equiv \ln \mathbf{M}_0(L)/L, \quad \mathbf{F}(z) \equiv \mathbf{M}_0(z)\exp(-\mathbf{K}z), \qquad (12)$$

and in view of $\mathbf{M}_0^{-1}(L) = \exp(-\mathbf{K}L)$ and (11), we deduce the equality
$$\mathbf{F}(z + L) = \mathbf{M}_0(z + L)\exp[-\mathbf{K}(z + L)] = \mathbf{M}_0(z)\exp(-\mathbf{K}z) = \mathbf{F}(z). \qquad (13)$$

Thus the matrix $\mathbf{F}(z)$ is a periodic one. By using (8), the transfer matrix of the Cauchy problem for equation (5) with periodic coefficient is derived like this

$$\mathbf{M}(z, z_0) = \mathbf{F}(z)\exp[\mathbf{K}(z - z_0)]\mathbf{F}^{-1}(z_0). \qquad (14)$$

In this case the corresponding general solution of (5) has a form

$$\mathbf{\Psi}(z) = \mathbf{F}(z)\exp(\mathbf{K}z)\mathbf{c} = \mathbf{F}(z)\mathbf{c}\exp(\mathbf{c}^{-1}\mathbf{K}\mathbf{c}z), \qquad (15)$$

where $\mathbf{c}$ is some arbitrary constant matrix compiled from the set of fundamental solutions of equation (5) in $z = 0$, $\det \mathbf{c} \neq 0$. Note, $\mathbf{c}$ can be linked to $\mathbf{\Psi}(z)$ via the substitutions $\tilde{\mathbf{F}}(z) = \mathbf{F}(z)\mathbf{c}$, $\tilde{\mathbf{K}} = \mathbf{c}^{-1}\mathbf{K}\mathbf{c}$. The formula (15) is a vector analogue of the Floquet theorem [19]. The use of the Floquet's theorem for solutions of the system (5) with periodical coefficients can be more effective than a direct numerical solution. Actually, according to (15), in order to find a solution suitable for any $z$, it suffices to define the function $\mathbf{F}$ in one period and to find the constant matrix $\mathbf{K}$. Both latter tasks are provided by the knowledge of the matrix $\mathbf{M}_0(z)$ in one period $0 \leq z \leq L$; which is sufficient to solve the problem for $z_0 = 0$ with the initial condition $\mathbf{M}(z_0, z_0) = \mathbf{I}$ on the interval $[0, L]$.

In our case the structure period consists of two layers with thicknesses $d_1$ and $d_2$ (Fig. 1). Therefore the system is described by the transfer matrices $\mathbf{M}_1 = \mathbf{M}(d_1, 0)$ and $\mathbf{M}_2 = \mathbf{M}(L, d_1)$ related to intervals $0 \leq z \leq d_1$ and $d_1 \leq z \leq L$ respectively. As follows from expression (6) and relations (7), the transfer matrix of the structure period is a product of the transfer matrices of layers $\mathbb{M} = \mathbf{M}_2\mathbf{M}_1$ and the field components referred to boundaries of the first double-layer period of the structure are related as

$$\mathbf{\Psi}(L) = \mathbf{M}_2\mathbf{\Psi}(d_1) = \mathbf{M}_2\mathbf{M}_1\mathbf{\Psi}(0) = \mathbb{M}\mathbf{\Psi}(0). \qquad (16)$$

Note that $\mathbb{M}$ is the monodromy matrix related to equation (5) with periodical coefficients in a special case of double-layer structure period.

It is obvious, that the field components referred to the outer boundaries of the whole structure consisted of $N$ periodic slabs are connected by transfer matrix $\mathbb{M}$ raised to the power $N$

$$\mathbf{\Psi}(NL) = \mathbb{M}^N\mathbf{\Psi}(0). \qquad (17)$$

In order to search the fields reflected from and transmitted through structures with a large number of periods $N \gg 1$, for the raising of a matrix $\mathbb{M}$ to power $N$, the algorithm of the matrix polynomial theory can be used [21, 22]

$$\mathbb{M}^N = \sum_{j=1}^{4} \rho_j^N \mathbf{P}_j, \quad \mathbf{P}_j = \mathbf{V}\mathbf{E}_j\mathbf{V}^{-1}, \qquad (18)$$

where $\rho_j$ are the eigenvalues of the transfer-matrix $\mathbb{M}$, $\mathbf{V}$ is a matrix whose columns are the set of independent eigenvectors of $\mathbb{M}$, $\mathbf{E}_j$ is the matrix with 1 in the $(j, j)$-locations and zeros elsewhere.

## 4. Bloch eigenwaves theory

One of the approaches based on the Floquet's theorem (15), is the method of Bloch waves that describes the propagation conditions of eigenwaves of an infinite periodic structure. It is based on the fact that for any *multiplicator* $\rho$ there is a nontrivial solution of the periodic system (5), satisfying the condition [19]

$$\mathbf{\Psi}(z + L) = \rho\mathbf{\Psi}(z), \qquad (19)$$

where multiplicators are the eigenvalues of the monodromy matrix, i.e. they are the roots of the following characteristic polynomial

$$\det[\mathbb{M} - \rho\mathbf{I}] = 0. \qquad (20)$$

As follows from (12), the eigenvalues $\gamma$ of the matrix $\mathbb{K} = \ln\mathbb{M}/L$ satisfy the equation

$$\det[\mathbb{K} - \gamma\mathbf{I}] = 0 \qquad (21)$$

and the multiplicators $\rho$ are related via the condition

$$\gamma_j = \frac{1}{L}\ln\rho_j = \frac{1}{L}\Big[\ln|\rho_j| + \mathrm{i}(\arg\rho_j + 2n\pi)\Big], \quad (j=\overline{1,4},\ n=0,\pm1,\pm2,\ldots). \tag{22}$$

In our case, $\mathbf{A}(z)$ is the $4\times 4$ matrix that satisfies the condition $\mathrm{tr}[\mathbf{A}(z)] = 0$, where tr denotes the matrix trace, i.e. the sum of all elements of matrix main diagonal. Therefore, according to the Liouville-Jacobi formula [19]

$$\det\mathbb{M} = \exp\int_0^L \mathrm{tr}[\mathbf{A}(t)]dt = 1. \tag{23}$$

Thus the matrix $\mathbb{M}$ is unimodular, and, after the determinant calculating, equation (20) comes to the following polynomial form

$$\rho^4 + S_3\rho^3 + S_2\rho^2 + S_1\rho + S_0 = 0, \tag{24}$$

where $S_0 = \det\mathbb{M} = 1$, $S_1 = -\sum_{i=1}^{2}\sum_{j=i+1}^{3}\sum_{k=j+1}^{4}\big(m_{ii}m_{jj}m_{kk} + m_{ij}m_{jk}m_{ki} + m_{ik}m_{ji}m_{kj} - m_{ii}m_{jk}m_{kj} - m_{jj}m_{ki}m_{ik} - m_{kk}m_{ij}m_{ji}\big)$, $S_2 = \sum_{i=1}^{3}\sum_{j=i+1}^{4}(m_{ii}m_{jj} - m_{ij}m_{ji})$, $S_3 = -\sum_{i=1}^{4}m_{ii}$, and $m_{\alpha\beta}$ are the elements of the matrix $\mathbb{M}$. If coefficients of equation (24) are satisfy to equality $S_3 = S_1$, the left part of the dispersion equation of this type can be represented as the product of two quadratic polynomials [16, 17, 23]

$$\big[\rho^2 + Q_1\rho + 1\big]\big[\rho^2 + Q_2\rho + 1\big] = 0. \tag{25}$$

The fact that the condition $S_3 = S_1$ is satisfied can be verified numerically. In this case the coefficients of the equations (24) and (25) are related as the next

$$Q_1 + Q_2 = S_1, \quad 2 + Q_1Q_2 = S_2. \tag{26}$$

Expressing $Q_1$ and $Q_2$ through the $S_1$ and $S_2$, we obtain

$$Q_{1,2} = \frac{S_1}{2} \pm \sqrt{\left(\frac{S_1}{2}\right)^2 + 2 - S_2}. \tag{27}$$

Thus, the dispersion equation (25) is split into two independent parts. From a physical point of view it means that in the structure, there are two independent spectra of eigenwaves, each of them is characterized by its dispersion relation and wavenumber.

From (27) the eigenvalues of the transfer matrix of one period $\mathbb{M}$ can be written in the form

$$\rho_{1,2} = -\frac{S_1}{4} - \sqrt{\left(\frac{S_1}{4}\right)^2 + \frac{2-S_2}{4}} \pm \sqrt{\frac{1}{4}\left[\frac{S_1}{2} + \sqrt{\left(\frac{S_1}{2}\right)^2 + 2 - S_2}\right]^2 - 1},$$

$$\rho_{3,4} = -\frac{S_1}{4} + \sqrt{\left(\frac{S_1}{4}\right)^2 + \frac{2-S_2}{4}} \pm \sqrt{\frac{1}{4}\left[\frac{S_1}{2} - \sqrt{\left(\frac{S_1}{2}\right)^2 + 2 - S_2}\right]^2 - 1}. \tag{28}$$

The multiplicators are related to the propagation constants of the eigenwaves via the condition $\rho_j^{\pm} = \exp(\pm\mathrm{i}\gamma_j L)$; the sign choice for the $j$-th type of wave corresponds to the wave propagation direction.

To analyze the stability of the obtained solutions, according to the Lyapunov theorem on the *reducibility* [19], the changing of variables in equation (5)

$$\mathbf{\Psi} = \mathbf{F}(z)\mathbf{\Upsilon} = \mathbf{M}_0(z)\exp(-\mathbf{K}z)\mathbf{\Upsilon} \tag{29}$$

leads equation (5) with the periodic matrix $\mathbf{A}(z)$ to the equation with the constant matrix $\mathbf{K}$

$$\frac{\partial}{\partial z}\mathbf{\Upsilon} = \mathbf{K}\mathbf{\Upsilon}. \tag{30}$$

The general solution of this equation, as known, is given by
$$\Upsilon = \mathbf{c}\exp(\mathbf{K}z). \tag{31}$$
From all the solutions of this equation we select the trivial solution $\Upsilon = 0$, and ask the question about of its stability, namely: do small deviations from this solution at $z = 0$ lead to small deviations for all $z \geq 0$? The answer varies depending on the form of the matrix $\mathbf{K}$. From the general definition of the *Lyapunov stability* it is implied that the solutions are stable if and only if the matrix $\exp(\mathbf{K}z)$ is bounded for all $z \geq 0$, and the solutions are asymptotically stable if the matrix $\exp(\mathbf{K}z)$ tends to zero for all $z \geq 0$. Thus the stability of the system (30) is completely determined by the form of the roots $\gamma_j$ of the characteristic polynomial of the matrix $\mathbf{K}$ (21), and these conditions are determined via the theorems of Lyapunov [19].

On the basis of this theorem, using the relation (22), the following conditions of the solution stability of the periodic system (5) can be formulated. The solutions are stable if all the multiplicators $\rho_j$ lie within the boundaries of the closed unit circle $|\rho_j| \leq 1$ ($|Q_j| \leq 2$). The multiplicators that lie on the circle $|\rho_j| = 1$ ($|Q_j| = 2$) possess simple elementary divisors. For the asymptotic stability of the solutions of the periodic system, it is necessary and sufficient that all the multiplicators lay inside the unit circle $|\rho_j| < 1$ ($|Q_j| < 2$).

From the electromagnetic point of view, when considering the periodic structure, the regions of stability and instability of solutions of equation (5) correspond to regions where waves propagate or do not propagate. Thus, in the first case $|Q_j| < 2$, the frequency range and basic-element parameters provide the propagation of the *j*-th wave (passbands). In the second case $|Q_j| > 2$, the wave does not propagate (stopbands), and $\rho_j^\pm \equiv \exp(\pm\gamma_j^{//}L)$. The band edges are the regimes where $|Q_j| = 2$ ($\gamma_j = 0$).

Note that in a periodic structure without losses, multiplicators are usually located on a circle of unit radius or on the real axis. In some anisotropic structures the points $|\rho_j| = 1$ can be located at arbitrary points of the unit circle on the complex plane of $\rho$. These degenerate points are of particular interest and are a subject of some special studies [24, 25] related to the problem of "slow light". The advantage of the general theoretical approach used here is its applicability for the analysis of fields in the periodic structures under a degeneracy of their eigenwaves.

## 5. Reflection and transmission coefficients

To find the reflection and transmission coefficients, we use the solution (17) of equation (5) that can be equivalently formulated as
$$\mathbf{\Psi}(0) = (\mathbb{M}^N)^{-1}\mathbf{\Psi}(NL) = \mathbb{T}\mathbf{\Psi}(NL). \tag{32}$$

The field vector at the input surface is made up of two parts that consist of the incident and reflected wave contributions
$$\mathbf{\Psi}(0) = \mathbf{\Psi}_{in} + \mathbf{\Psi}_{ref}. \tag{33}$$
The field at the output surface matches only a single transmitted wave field
$$\mathbf{\Psi}(NL) = \mathbf{\Psi}_{tr}. \tag{34}$$
On the other hand, the incident, reflected, and transmitted field can be written as follow
$$\vec{E}^{in}(\vec{r}) = \vec{E}_0^{in}\exp(i\vec{k}^{in}\cdot\vec{r}), \quad \vec{E}^{ref}(\vec{r}) = \vec{E}_0^{ref}\exp(i\vec{k}^{ref}\cdot\vec{r}), \quad \vec{E}^{tr}(\vec{r}) = \vec{E}_0^{tr}\exp(i\vec{k}^{tr}\cdot\vec{r}). \tag{35}$$
The fields (35) can be represented in terms of the linearly polarized waves. Thus the field components in the input and output half-spaces are (the factor $\exp[-i(\omega t - k_x x - k_y y)]$ is omitted) like this

$$\begin{Bmatrix} E_{x0} \\ E_{y0} \end{Bmatrix} = \pm \begin{Bmatrix} 1/\sqrt{Y_0^s} \\ 1/\sqrt{Y_0^p} \end{Bmatrix} \left( \begin{Bmatrix} A^s \\ A^p \end{Bmatrix} \exp(ik_{z0}z) \pm \begin{Bmatrix} B^s \\ B^p \end{Bmatrix} \exp(-ik_{z0}z) \right),$$

$$\begin{Bmatrix} H_{y0} \\ H_{x0} \end{Bmatrix} = \begin{Bmatrix} \sqrt{Y_0^s} \\ \sqrt{Y_0^p} \end{Bmatrix} \left( \begin{Bmatrix} A^s \\ A^p \end{Bmatrix} \exp(ik_{z0}z) \mp \begin{Bmatrix} B^s \\ B^p \end{Bmatrix} \exp(-ik_{z0}z) \right),$$

(36)

$$\begin{Bmatrix} E_{x3} \\ E_{y3} \end{Bmatrix} = \pm \begin{Bmatrix} \left(1/\sqrt{Y_3^s}\right)C^s \\ \left(1/\sqrt{Y_3^p}\right)C^p \end{Bmatrix} \exp[ik_{z3}(z-NL)], \quad \begin{Bmatrix} H_{y3} \\ H_{x3} \end{Bmatrix} = \begin{Bmatrix} \sqrt{Y_3^s}C^s \\ \sqrt{Y_3^p}C^p \end{Bmatrix} \exp[ik_{z3}(z-NL)]. \quad (37)$$

Here $A^v$, $B^v$ and $C^v$ ($v=p,s$) are the amplitudes of the incident, reflected and transmitted field, respectively; $Y_j^s = \eta_j^{-1}\cos\theta_j$, $Y_j^p = (\eta_j \cos\theta_j)^{-1}$ ($j=0,3$) are the wave admittances of input and output half-spaces; $k_{zj} = k_j \cos\theta_j$, $k_j = kn_j$, $n_j = \sqrt{\varepsilon_j\mu_j}$, $\eta_j = \sqrt{\mu_j/\varepsilon_j}$, $\sin\theta_j = \sin\theta_0 n_0/n_j$, and the term $s$ is related to the perpendicular polarization (electric-field vector $\vec{E}$ is perpendicular to the plane of incidence) and the term $p$ is related to the parallel polarization (electric-field vector $\vec{E}$ is parallel to the plane of incidence) of plane electromagnetic waves.

The substitution of (36), (37) at the interfaces $z=0$ and $z=NL$ into (33), (34) yields the following system of algebraic equations

$$A^s + B^s = \sqrt{\frac{Y_0^s}{Y_3^s}} a_1^s C^s + \sqrt{\frac{Y_0^s}{Y_3^p}} a_1^p C^p, \quad A^p - B^p = -\sqrt{\frac{Y_0^p}{Y_3^s}} a_2^s C^s - \sqrt{\frac{Y_0^p}{Y_3^p}} a_2^p C^p,$$

$$A^s - B^s = \frac{a_4^s}{\sqrt{Y_0^s Y_3^s}} C^s + \frac{a_4^p}{\sqrt{Y_0^p Y_3^s}} C^p, \quad A^p + B^p = \frac{a_3^s}{\sqrt{Y_0^p Y_3^s}} C^s + \frac{a_3^p}{\sqrt{Y_0^p Y_3^p}} C^p,$$

(38)

where $a_j^s = t_{j1} + t_{j4}Y_3^s$, $a_j^p = -t_{j2} + t_{j3}Y_3^p$, $j=\overline{1,4}$, and $t_{\alpha\beta}$ are the elements of the transfer matrix $\mathbb{T}$.

Then, we assume that incident field is either $p$-type ($A^s = 0$), either $s$-type ($A^p = 0$), and the co-polarized reflection and transmission coefficients are determined by the expressions $R^{vv} = B^v/A^v$ and $\tau^{vv} = C^v/A^v$, and the cross-polarized ones are $R^{vv'} = B^{v'}/A^v$ and $\tau^{vv'} = C^{v'}/A^v$, respectively. From (38) they are:

$$R^{ss} = \left(b_{sp}^- b_{ps}^- - b_{ss}^- b_{pp}^-\right)/\Delta, \qquad \tau^{ss} = -2\sqrt{Y_0^s Y_3^s}\, b_{ss}^-/\Delta,$$

$$R^{sp} = 2\sqrt{Y_0^s Y_0^p}\left(a_2^p b_{sp}^- - a_2^s b_{ss}^-\right)/\Delta, \quad \tau^{sp} = 2\sqrt{Y_0^s Y_3^p}\, b_{sp}^-/\Delta,$$

$$R^{pp} = \left(b_{pp}^+ b_{ss}^+ - b_{ps}^+ b_{sp}^+\right)/\Delta, \qquad \tau^{pp} = 2\sqrt{Y_0^p Y_3^p}\, b_{pp}^+/\Delta,$$

$$R^{ps} = 2\sqrt{Y_0^p Y_0^s}\left(a_1^p b_{pp}^+ - a_1^s b_{ps}^+\right)/\Delta, \quad \tau^{ps} = -2\sqrt{Y_0^p Y_3^s}\, b_{ps}^+/\Delta,$$

(39)

where $\Delta = b_{ps}^+ b_{sp}^- - b_{pp}^+ b_{ss}^-$, and $b_{pp}^{\pm} = Y_0^p a_2^p \pm a_3^p$, $b_{ps}^{\pm} = Y_0^p a_2^s \pm a_3^s$, $b_{ss}^{\pm} = Y_0^s a_1^s \pm a_4^s$, $b_{sp}^{\pm} = Y_0^s a_1^p \pm a_4^p$.

The polarization state of both reflected and transmitted fields can be obtained using standard definition [26]: $\tan 2\beta = U_2/U_1$, $\sin 2\eta = U_3/U_0$, where $\beta$ is the polarization azimuth, $\eta$ is the ellipticity angle (Fig. 3). $U_j$ are the Stokes parameters calculated from the components of the electric field in the right-handed orthogonal frame.

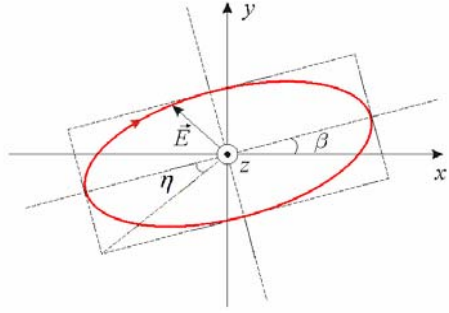

**Figure 3. (color online)** Parameters of the polarization ellipse.

### 6. Numerical results and discussion.

At first we consider the case of normal incidence ($\theta_0 = 0$) of a plane wave on a MPLS. For the sake of simplicity, but without loss of generality, let us assume that the structure consists of ferrite layers separated by air gaps, i.e. $\varepsilon_2 = \mu_2 = 1$ is assigned.

As discussed above, the stability and instability domains of solutions of the system (5) correspond to the regions of problem parameters in which propagation of electromagnetic waves is permitted or forbidden respectively. Thus the values of $Q_{1,2}$ defined by expression (27) determine the band spectrum of two pairs of eigenwaves. One pair of eigenwaves propagates in direction of z-axis and the other has the opposite direction of propagation. The passbands are determined by the condition $|Q_1| \leq 2$ for one eigenwave and $|Q_2| \leq 2$ for another in each of these pairs (Fig. 4). These conditions are displayed in Fig. 4a as a shaded area. Since the dispersion equation (25) consists of two independent factors, the bandwidth of these spectra can be mutually overlapped. One can see a significant difference between these two solutions of the dispersion equation. This is because they correspond to two waves with different polarizations.

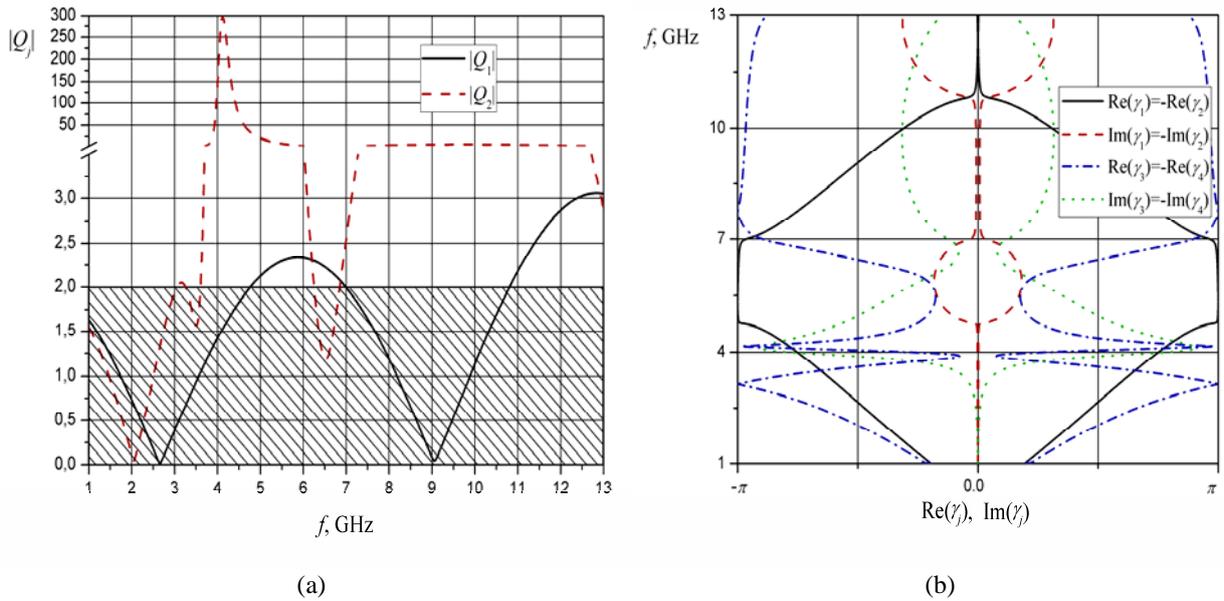

(a)           (b)

**Figure 4. (color online)** Stability conditions (a) and band spectrum (b) of magnetophotonic layered structure under normal wave propagation ($\theta_0 = 0$), $d_1 = d_2 = 5 \times 10^{-3}$ m, $\varphi_0 = 0$.

As is well known, the eigenwaves of an unbounded ferrite medium propagating along the magnetization direction are the left-handed (LCP) and right-handed (RCP) circularly polarized waves which differ in the propagation constants $\Gamma^{\pm} = k\sqrt{\varepsilon(\mu^T \pm \alpha)} = k\sqrt{\varepsilon\mu^{\pm}}$ [7, 8]. Thus each of those waves propagates in the medium with different effective magnetic permeability $\mu^{\pm}$, where $\mu^+$ and $\mu^-$ are related to the LCP and RCP waves, respectively. For clarity, the frequency dependences of $\mu^{\pm}$ are shown in Fig. 5. In our case, $\text{Re}(\mu^-)$ is a positive value over the chosen frequency range, and the medium losses are infinitesimal for the RCP wave ($\text{Im}(\mu^-) \sim 0.01$). On the other hand, it is possible to select three specific frequency ranges where $\mu^+$ acquires different properties. In the first range, located between 1 GHz and 3 GHz, the effective permeability $\mu^+$ has a positive value of real part and a small imaginary part. In the second range, between 3 GHz and 5 GHz the real part of $\mu^+$ varies from positive values to negative ones as frequency increases. This transition takes a place at the frequency of the ferromagnetic resonance ($f_0 = 4$ GHz). In this range the medium losses are very significant. Finally, in the third frequency range, located from 5 GHz to 10 GHz, the permeability $\mu^+$ has a negative real part and a small imaginary part.

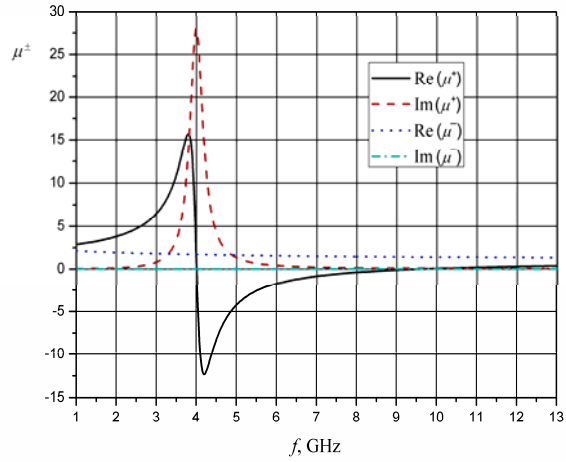

**Figure 5. (color online)** Frequency dependences of effective permeability $\mu^{\pm}$ related to the left-handed and right-handed circularly polarized waves.

The properties of the ferrite medium affect the propagation conditions of the eigenwaves of the periodic structure. On the basis of equations (22) and (28), the propagation constants $\gamma_j$ ($j = \overline{1,4}$) of the eigenwaves are obtained via the solutions of the dispersion equations (27) with $Q_1$ and $Q_2$. These propagation constants obey the conditions $\gamma_1 = -\gamma_2$ and $\gamma_3 = -\gamma_4$, where the sign defines the propagation direction along or opposite to z-axis. Thus, it is defined that $\gamma_1$, $\gamma_3$ and $\gamma_2$, $\gamma_4$ correspond to the RCP waves and LCP waves, respectively. In the case of the RCP wave propagating in the positive z-axis direction, due to small dissipative losses related to $\mu^-$, the eigenwave spectrum has interleaved passbands given by $|Q_1| < 2$, $\text{Im}\gamma_1 = \text{Im}\gamma_2 \approx 0$ and stopbands given by $|Q_1| > 2$, $\text{Im}\gamma_1 = -\text{Im}\gamma_2 \neq 0$. The band edges are the regimes that correspond to $|Q_1| = 2$. In the case of the LCP wave, starting with some frequency near 3 GHz, it is no longer possible to select the alternation of the passband and stopband positions. As is clear from Fig. 4, the imaginary parts of $\gamma_3$ and $\gamma_4$ are significant above this frequency, and the condition $|Q_2| > 2$ holds almost over all selected frequency range.

The mentioned features of the eigenwaves of the infinite periodic structure composed of ferrite layers appear particularly in the frequency dependences of the reflection and transmission coefficients of the LCP and RCP waves of the bounded stack formed of $N$ basic double-layer elements (see Fig. 6). These frequency dependences have interleaved bands of the reflection and transmission that corresponds to the stopbands and passbands of the eigenwaves. The finite number of the structure periods results as small-scale oscillations in the passbands. These oscillations are a consequence of the interference with waves reflected from outside boundaries of the stack. The number of oscillations is $N-1$ within each passband. Evidently, the dissipative losses reduce the average level of the reflection and transmission in the whole frequency range and decrease the amplitude of the small-scale interference oscillations in the passbands. Also note that a finite number of stack slabs results in a partial transmission of the waves within the stopbands.

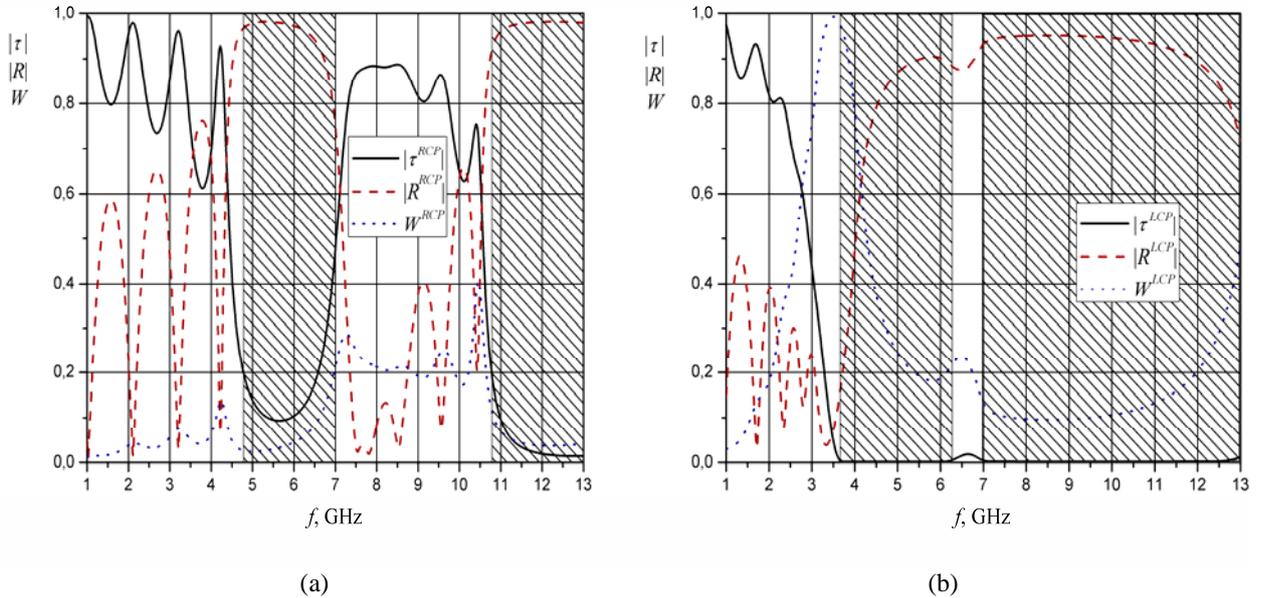

(a)            (b)

**Figure 6. (color online)** Frequency dependences of transmission, reflection and absorption $W = 1 - |R^\nu|^2 - |\tau^\nu|^2$ coefficients of right-handed ($\nu = \text{RCP}$) (a) and left-handed ($\nu = \text{LCP}$) (b) circularly polarized waves under normal incidence ($\theta_0 = 0$), $N = 5$, $d_1 = d_2 = 5 \times 10^{-3}$ m, $\varphi_0 = 0$.

All these properties are typical for both the LCP and RCP waves. But for the LCP wave some distinctive features can be pointed out. In the frequency band near the frequency of ferromagnetic resonance $\omega_0$ almost all the energy of the LCP wave is absorbed. Note that this effect of the different absorption of LCP and RCP waves is well known in optics as circular dichroism. At the higher frequency, the real part of ferrite effective permeability $\mu^+$ is negative, which leads to a great imaginary value of the propagation constant $\Gamma^+$. In this band the LCP wave is reflected except for a portion of the absorbed energy.

Analyzing the properties of the reflected and transmitted fields in terms of a linearly polarized wave (Fig. 7), we note that generally a transformation of a linearly polarized wave to an elliptically polarized one appears at the MPLS output, and, over the whole selected frequency range the conditions $|\tau^{ss}| = |\tau^{pp}|$, $|\tau^{sp}| = |\tau^{ps}|$ and $|R^{ss}| = |R^{pp}|$, $|R^{sp}| = |R^{ps}|$ are satisfied under normal wave incidence. The above mentioned peculiarities of the absorption and reflection of the LCP wave lead the degeneration of the elliptical polarization to the circular polarization of the transmitted field, i.e. the magnitudes of the co-polarized and cross-polarized components of the transmitted field are equal to each other, $|\tau^{ss}|=|\tau^{pp}|=|\tau^{sp}|=|\tau^{ps}|$, and the ellipticity angle is $\eta=-\pi/4$. This condition is observed in the frequency range starting from the frequency of a ferromagnetic resonance and up to the frequency where $\mu^+$ becomes positive. At some frequencies the elliptical polarization changes to a linear polarization, which corresponds to $\eta=0$. For the reflected field this situation appears at the frequencies of the band edges. Within the passbands the conditions are also possible when both the reflected and transmitted fields are circularly polarized. One of these cases is shown in Figs. 7c, 7d by dotted lines which corresponds to the frequency at 7.8 GHz.

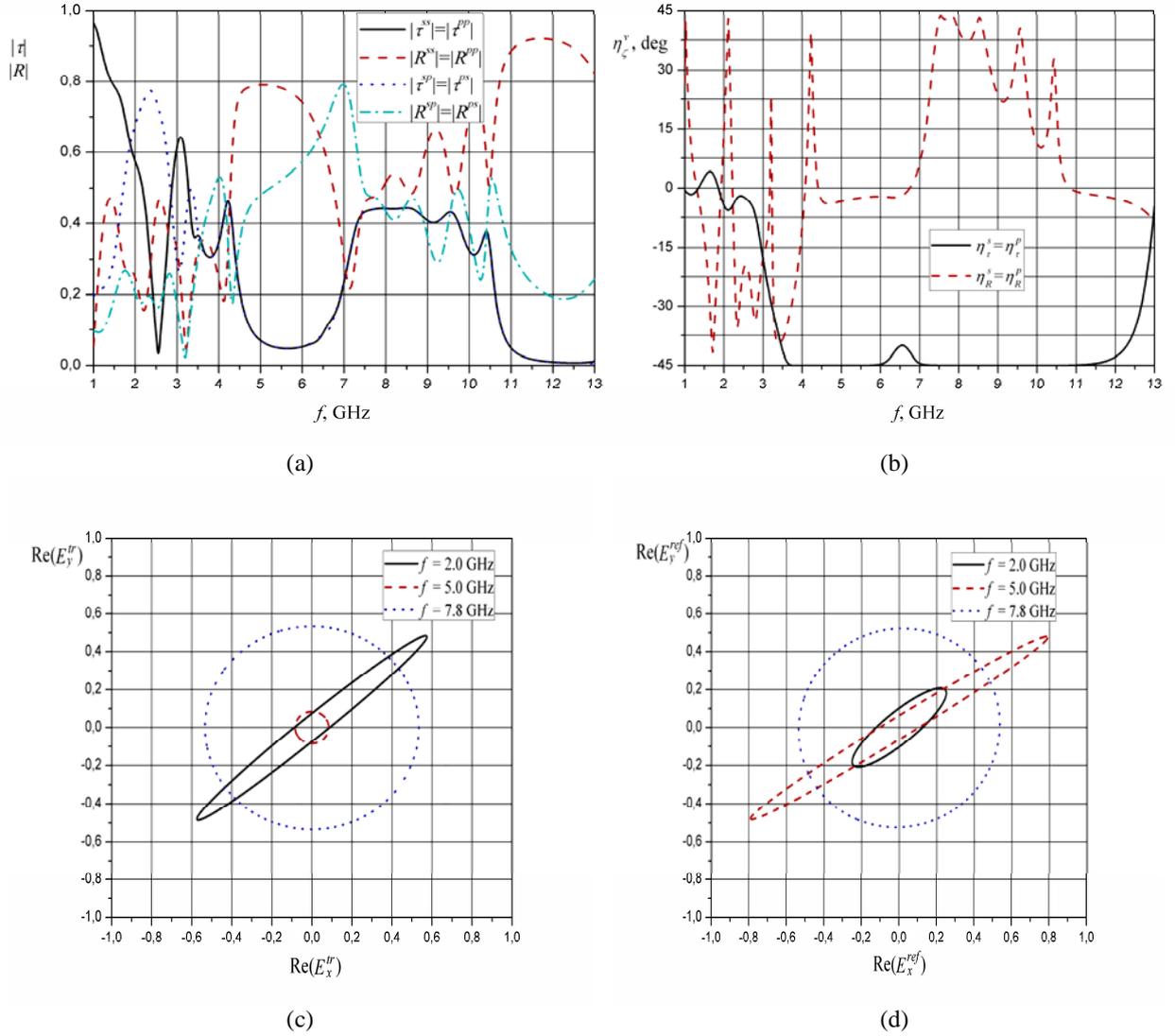

**Figure 7. (color online)** Frequency dependences of reflection and transmission coefficients (a), ellipticity (b) and polarization ellipse of transmitted (c) and reflected (d) fields under normal incidence ($\theta_0=0$) of the linearly $x$-polarized wave, $N=5$, $d_1=d_2=5\times10^{-3}$ m, $\varphi_0=0$.

It is interesting to further investigate the influence of the structure periodicity on the enhancement of the Faraday rotation as a result of the wave interference within the multilayer stack (see Fig. 8). For this study three different structure compositions are considered. In the first case the structure consists of a single homogeneous ferrite layer with finite thickness. In the second structure configuration, there are two ferrite layers separated by an air gap. And in the third case, the structure is a periodical stack of four ferrite layers separated by air gaps. In general, the total thickness of the ferrite layers in all the structure configurations remains unchanged.

In terms of the linearly polarized wave the enhancement of the Faraday rotation can be estimated through the level of the amplitude of the co-polarized and cross-polarized components of the transmission and reflection coefficients, since these components are directly related to the definition of the polarization ellipse (Fig. 3). As is seen in Fig. 8a, for the transmitted field, the enhancement of the Faraday effect is observed in case when both the permeabilities $\mu^+$ and $\mu^-$ are positive values. As pointed out above, under normal wave incidence and in the frequency range where $\mu^+$ is negative, the transmitted field is circularly polarized and the structure configuration difference occurs only in the level of the transmission coefficient magnitude.

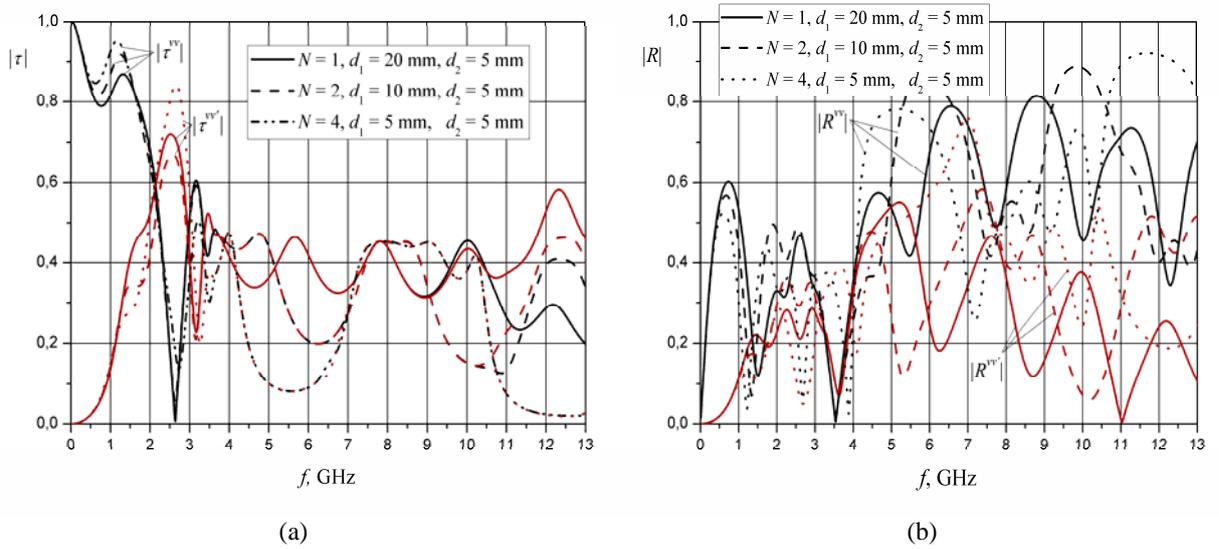

(a) (b)

**Figure 8. (color online)** Enhancement of Faraday rotation depending on the thicknesses and number of ferrite layers for transmitted (a) and reflected (b) fields, $\theta_0 = \varphi_0 = 0$.

If we consider the properties of the reflected field, the fact of the strong influence of the structure configuration on the magnitude of the co-polarized and cross-polarized reflection coefficients can be established (Fig. 8b). Thus, a choice of thicknesses and number of ferrite layers of the structure can produce either enhancement or weakening of the Faraday rotation in the reflected field at the given frequency.

The angle of incidence of the primary field significantly affects the relation between the level of the co-polarized and cross-polarized reflection and transmission. Three pairs of graphs are presented in Figure 8 to show the magnitudes of the reflected and transmitted fields. These pairs are related to the different frequencies that correspond to the different values of $\mu^+$.

At the frequency 2 GHz (Figs. 9a, 9b), the ferrite effective permeability $\mu^+$ is positive and magnetic losses are vanishingly small. This frequency corresponds to the wavelength that is much greater than the structure period length and the MPLS can be considered as a homogeneous gyrotropic layer of the same length $NL$. It is clear that the angular dependences of the transmission and reflection coefficient represent this situation. Thus there are no significant variations of the magnitudes of the reflection and transmission coefficients, because the interference of waves inside the structure does not occur. Note that the condition $|R^{ps}|=|R^{sp}|$ is satisfied for all angles of incidence but for a certain angles $|\tau^{ps}|\neq|\tau^{sp}|$.

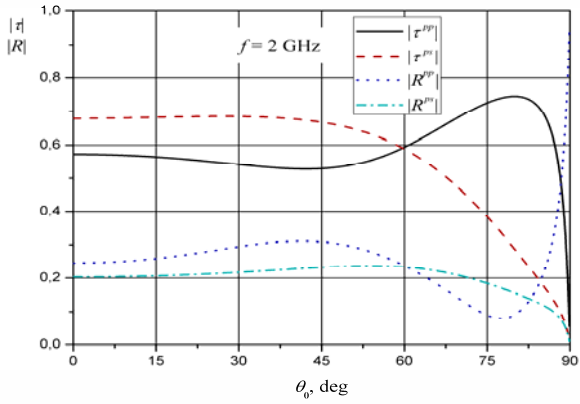
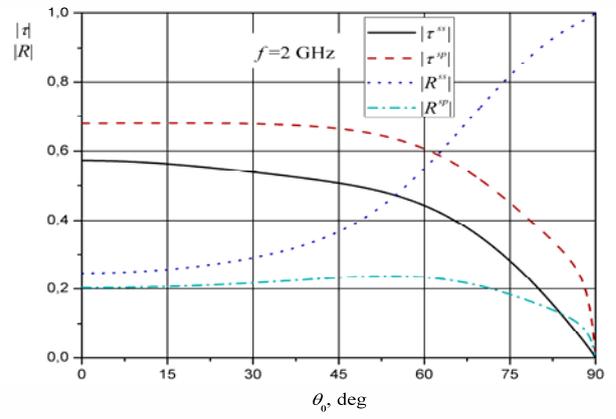

(a)                                                      (b)

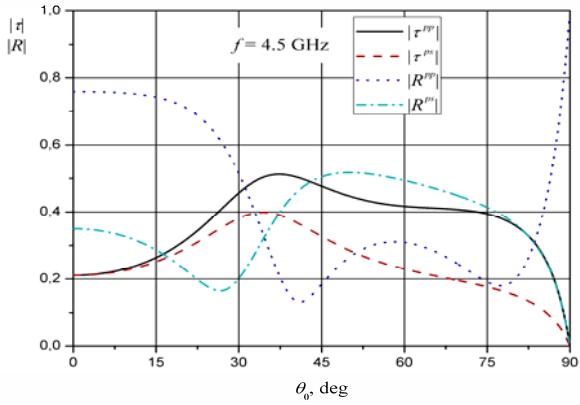
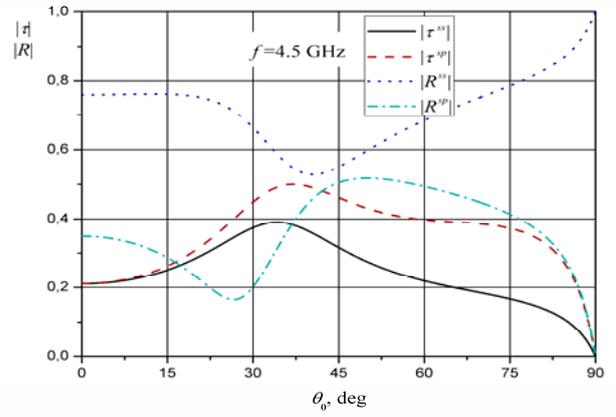

(c)                                                    (d)

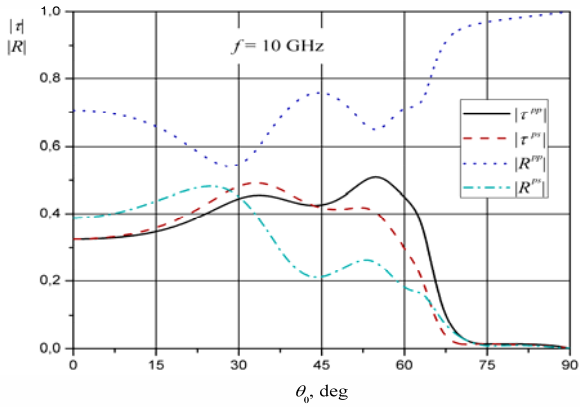
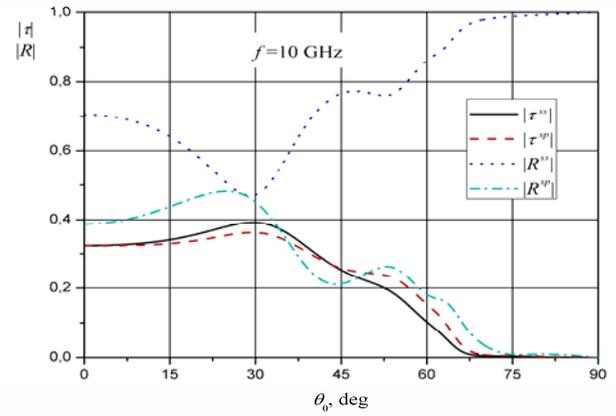

(e)                                                    (f)

**Figure 9. (color online)** Angular dependences of reflection and transmission coefficients of linearly polarized wave; $N = 5$, $d_1 = d_2 = 5 \times 10^{-3} m$, $\varphi_0 = 0$.

The frequency 4.5 GHz is close to the frequency of ferromagnetic resonance (Figs. 9c, 9d). Here $\mu^+$ is negative and the magnetic losses are significant. As mentioned above, under the normal incidence ($\theta_0 = 0$), the transmitted field is circularly polarized. Oblique incidence leads to the elliptical polarization of the transmitted field. The ellipses related to orthogonally polarized waves have identical ellipticity parameters defined by the following conditions $|\tau^{pp}|=|\tau^{sp}|$, $|\tau^{ss}|=|\tau^{ps}|$.

The last pair of the graphs corresponds to the frequency of 10 GHz (Figs. 9e, 9f). At this frequency $\mu^+$ is close to zero and magnetic losses are vanishingly small. In this case, starting from a certain cutoff angle ($\theta_0 \approx 65$, degrees), the wave is completely reflected from the MPLS. The polarization of this reflected field is linear and similar to the polarization of the primary incident field, i.e. $|R^{ps}|=|R^{sp}|=0$. Note that at this frequency for all angles of incidence, the components of the reflected field are practically equal to each other for the orthogonally polarized waves, i.e. the conditions $|R^{pp}|\approx|R^{ss}|$, $|R^{ps}|=|R^{sp}|$ are satisfied.

**7. Conclusion**

In conclusion, we studied the electromagnetic properties of a magnetophotonic layered structure at the frequencies close to the frequency of ferromagnetic resonance. The research was carried out taking into account a great value of dissipative losses in the ferrite layers of the structure in the frequency range. This circumstance requires generalizing the definition of eigenwaves of the periodic structure. It was stated that the eigenwaves are orthogonal in the energy sense and every eigenwave propagates independently of the others in the region which is free of sources.

The method of our study is based on deriving a system of ordinary differential equations on transversal field components and further analysis of the solutions stability of this system using the general theory of differential equations. Through the analysis of the eigenvalues of the transfer matrix of the structure period, the boundaries of the stopbands and passbands of the eigenwaves were determined since these eigenvalues describe the stability of the system solutions. We estimated the difference of the two kinds of eigenwaves in relation to the stability conditions of the system solution. In the assumption of propagation normal to structure layers, these two kinds of eigenwaves are the RCP and LCP waves. The particular absorption and reflection of the LCP wave was studied.

A dramatic effect of the biased ferrite material on the LCP wave propagation results in the polarization transformations of a linearly polarized plane wave impinged on the MPLS. Generally the reflected and transmitted fields have elliptical polarization. We ascertained that, under normal incidence, the transmitted field has circular polarization in the frequency range of negative values of the biased ferrite effective permeability related to the LCP wave. It is shown that in this frequency band, starting from some angle of incidence, the transmitted field becomes elliptically polarized. At the frequencies of the stopband edges, the reflected field is linearly polarized.

We observed an enhancement of the Faraday rotation by the periodic stack in comparison with the rotation by some effective ferrite slab. The polarization rotation was examined for different configurations of the stack. We have shown that the rotation enhancement takes place in the transmitted field at the frequencies corresponding to positive effective permeability of the ferrite for both RCP and LCP waves.

**References**
1. Yablonovitch E 1993 *J. Opt. Soc. Am. B.* **10** 283
2. Sakoda K 2001 *Optical Properties of Photonic Crystals* (Springer)
3. Yablonovitch E 2007 Optics & Photonics News **18** 12
4. Inoue M, Arai K, Fujii T and Abe M 1998 *J. Appl. Phys.* **83** 6768
5. Inoue M, Arai K, Fujii T and Abe M 1999 *J. Appl. Phys.* **85** 5768
6. Sakaguchi S and Sugimoto N 1999 *J. Opt. Soc. Am. A* **16** 2045

7. Mikaelyan A L 1963 *Theory and Application of Ferrites at Microwave Frequencies* (Gosenergoizdat, Moscow-Leningrad) [in Russian]
8. Gurevich A G 1963 *Ferrites at Microwave Frequencies* (Heywood, London)
9. Kato H, Matsushita T, Takayama A, Egawa M, Nishimura K and Inoue M 2003 *Opt. Commun*. **219** 271
10. Lyubchanskii I L, Dadoenkova N N, Lyubchanskii M I, Shapovalov E A and Rasing Th 2003 *J. Phys. D: Appl. Phys*. **36** R277
11. Inoue M, Fujikawa R, Baryshev A, Khanikaev A, Lim P B, Uchida H, Aktsipetrov O, Fedyanin A, Murzina T and Granovsky A 2006 *J. Phys. D: Appl. Phys*. **39** R151
12. Chernovtsev S V, Belozorov D P and Tarapov S I 2007 *J. Phys. D: Appl. Phys*. **40** 295.
13. Levy M and Jalali A A 2007 *J. Opt. Soc. Am. B* **24** 1603
14. Belozorov D P, Khodzitsky M K and Tarapov S I 2009 *J. Phys. D: Appl. Phys*. **42** 055003
15. Berreman D W 1972 *J. Opt. Soc. Am*. **62** 502
16. Bulgakov A A and Kononenko V K 2001 *Telecommunications and Radio Engineering* **55** 369
17. Vytovtov K A and Bulgakov A A 2006 *Telecommunications and Radio Engineering* **65** 1307
18. Il′inskiy A S and Slepyan G Ya 1983 *Oscillations and Waves in Electrodynamic Systems with Losses* (Moscow University Press, Moscow) [in Russian]
19. Jakubovich V A and Starzhinskij V H 1975 *Linear Differential Equations with Periodic Coefficients* (Wiley, New York), Vol. 1
20. Collin R E 1992 *Foundations for Microwave Engineering* (Wiley-Interscience, New York)
21. Dickey L J 1987 *ACM SIGAPL APL Quote Quad*. **18** 96
22. Tuz V R and Kazanskiy V B 2009 *J. Opt. Soc. Am. A* **26** 815
23. Tuz V R and Kazanskiy V B 2008 *J. Opt. Soc. Am*. *A* **25** 2704
24. Figotin A and Vitebsky I 2006 *Waves Rand. Compl. Media* **16** 293
25. Figotin A and Vitebsky I 2006 *Phys. Rev. E* **74** 066613
26. Collett E 1993 *Polarized Light: Fundamentals and Applications* (New York: Dekker)